\documentclass[oneside,english]{amsart}
\usepackage[OT1]{fontenc}
\usepackage{amsthm,amsmath}
\usepackage{natbib}
\usepackage[colorlinks,citecolor=blue,urlcolor=blue]{hyperref}

\usepackage[T1]{fontenc}
\usepackage[latin9]{inputenc}
\usepackage{float}
\usepackage{amsthm}
\usepackage{amssymb}
\usepackage{esint}

\makeatletter

\providecommand{\tabularnewline}{\\}

\numberwithin{equation}{section}
\numberwithin{figure}{section}

\usepackage{babel}

\makeatother

\usepackage{babel}
\begin{document}

\title{Robust estimation of isoform expression with RNA-Seq data}


\author[Jun Li]{Jun Li}
\address{Department of Applied and Computational \\
	Mathematics and Statistics\\
	153 Hurley Hall\\
	University of Notre Dame\\
	Notre Dame, IN 46556\\}
\email{jun.li@nd.edu}

\author[Hui Jiang]{Hui Jiang}
\address{Department of Biostatistics\\
	\&\\
	Center for Computational Medicine and Bioinformatics\\
	University of Michigan\\
	Ann Arbor, MI 48109}
\email{jianghui@umich.edu}

\begin{abstract}
Qualifying gene and isoform expression is one of the primary tasks
for RNA-Seq experiments. Given a sequence of counts representing numbers
of reads mapped to different positions (exons and junctions) of isoforms,
methods based on Poisson generalized linear models (GLM) with the
identity link function have been proposed to estimate isoform expression
levels from these counts. These Poisson based models have very limited
ability in handling the overdispersion in the counts brought by various
sources, and some of them are not robust to outliers. We propose a
negative binomial based GLM with identity link, and use a set of robustified
quasi-likelihood equations to make it resistant to outliers. An efficient
and reliable numeric algorithm has been identified to solve these
equations. In simulations, we find that our approach seems to outperform
existing approaches. We also find evidence supporting this conclusion
in real RNA-Seq data. 
\end{abstract}


\maketitle

\section{Introduction}

Through a regulated process called alternative splicing, most genes
in eukaryotes code for multiple types of mRNAs, which finally turn
into different proteins called ``isoforms''. As isoforms from the
same gene function differently, dysregulation of alternative splicing
can contribute to disease \citep[e.g.,]{lopez2005splicing}, and thus
it is of great importance and interest for biologists to study gene
expression at isoform level.

Before the appearance of the ultra-high-throughput sequencing (also
called the next-generation sequencing) technologies, genome-wide measurements
of gene expression mainly rely on microarrays, which have very limited
ability in discovering new isoforms and measuring isoform expression
as the design of microarrays relies on the reference genome/transcriptome.
In recent years, the ultra-high-throughput sequencing of transcriptomes
(RNA-Seq) is gradually taking place of microarrays and becoming arguably
the first choice in studying transcriptomes. A main advantage of RNA-Seq
is its ability in efficiently discovering new isoforms and studying
gene expression at isoform level. To quantitatively measure the expression
levels of isoforms, the number of reads (short sequences generated
by sequencing) mapped to each position of exons and junctions is counted.
In the ideal case, this number of reads can be modeled by a Poisson
distribution with mean being a linear combination of isoform expression
\citep{jiang2009statistical}. By maximizing the likelihood, algorithms
have been developed to estimate the isoform expression \citep{jiang2009statistical,trapnell2010transcript,li2011rsem}.
See \citet{pachter2011models} for a detailed review of the methods
for transcription quantification using RNA-Seq.

Although these methods have been quite successful, there are still
many challenges that seriously limit their reliability and accuracy.
Some of the challenges are recently pointed out by \citet{jiang2013penalized}:
systematic biases are often introduced during sequencing and mapping
processes, and the incompleteness in transcript annotation databases
also introduces additional uncertainly. To eliminate these effects,
they propose to add an L1-penalty term to the likelihood function
of the Poisson distribution. Their method has been shown to be able
to correct some of the biases in a robust manner.

In this paper, we argue that using Poisson distribution practically
limits the ability of the model to handle usually overdispersed read
counts, as well as various biases and uncertainties. We propose a
generalized linear model (GLM) based on negative binomial distributions
to efficiently handle these extra variation. Although negative binomial
based models with the log link function have been very popular for
the identification of differentially expressed genes based on RNA-Seq
data, such as edgeR \citep{edgeR}, DESeq \citep{DESeq}, ShrinkBayes
\citep{ShrinkBayes}, and baySeq \citep{baySeq}, they have not been
used for isoform expression estimation, and an important reason is
that isoform expression requires GLM with identity link, which automatically
brings in constraints in the parameter space, making it difficult
to solve the likelihood function. To circumvent this problem, previous
isoform expression estimation algorithms, which are based on Poisson
GLM with identity link, often uses Expectation-Maximization (EM) instead
of Newton-Raphson to give the estimate of parameters. However, such
a simple EM algorithm is not available for negative binomial distributions.

We get the solution of our model by solving a set of quasi-likelihood
equations. These equations use Huber-like penalties, so their solutions
are robust to outliers. Moreover, a simple (one-dimensional) primitive
function can be found for these functions, making the constraint optimization
convenient. On both simulated and real data, our method is able to
give more accurate and reliable estimate of isoform expression than
existing methods. To the best of our knowledge, this is the first
successful example of a robust negative binomial based GLM with the
identity link function.

\section{Robust quasi-likelihood equations for a negative-binomial regression
model}

\subsection{A negative-binomial regression model\label{sub:nb_model}}

In \citet{salzman2011statistical}, a Poisson regression model is
provided to model both single-end and paired-end RNA-Seq data for
isoform expression. We adopt their notations and extend it to a negative-binomial
regression model.

Suppose a gene has $I$ annotated distinct transcript isoforms and
$J$ possible distinct read types. Simply put, a read type is a group
of reads mapped to the same position of an exon or a junction. For
example, suppose a gene has only two isoforms; the first isoform has
only one exon of length 100 nt, and the second isoform is composed
of the exon and another exon of length 200 nt. Suppose each read is
single-end and of length 50 nt, then this gene has 251 possible distinct
read types: 51 types from exon 1 (positions 1 to 50 of exon 1, positions
2 to 51 of exon 1, $\ldots$, and positions 51 to 100 of exon 1),
151 types from exon 2, and 49 types from the junction (positions 52
to 100 of exon 1 + position 1 of exon 2, positions 53 to 100 of exon
1 + positions 1 to 2 of exon 2, $\ldots$, and position 100 of exon
1 + positions 1 to 49 of exon 2). We let $\theta$ be the $I\times1$
vector representing the abundance of the isoforms in the sample, and
$n$ be a $J\times1$ read count vector, where $n_{j}$ denotes the
number of reads of type $j$.

Previous methods assume the following Poisson distribution based model
\begin{equation}
n_{j}|\theta\sim\mbox{Poisson}\left(\sum_{i=1}^{I}\theta_{i}a_{ij}\right).\label{eq:poisson_regression}
\end{equation}
 On the above, $A=(a_{ij})$ is an $I\times J$ ``sampling rate''
matrix with its $(i,j)$-th element $a_{ij}$ denoting the rate that
read type $j$ is sampled from isoform $i$. Matrix $A$ describes
the compositions of the isoforms. In our previous example, $A$ will
be a $2\times251$ matrix, with the first 51 columns be $[1,1]$ and
the other 200 columns be $[0,1]$, meaning that the first 51 types
of reads can come from both isoforms and thus has a larger Poisson
mean $\theta_{1}+\theta_{2}$, while the other 200 types of reads
can only come from isoform 2 and thus has a smaller Poisson mean $\theta_{2}$.

In the example, all elements of matrix $A$ are either 0 or 1, showing
whether a read type can be generated from an isoform. In real data,
people have found that different read types, even from the same exon/junction,
can have quite different rates in sequencing \citep{Li10,Hansen10}.
For example, for read types 250 and 251 in our previous example, although
they are from the same exon and both come from isoform 2 for certain,
they still have means $a_{2,250}\theta_{2}$ and $a_{2,251}\theta_{2}$,
with $a_{2,250}\ne a_{2,251}$. These rates often depend on the nucleotide
composition of and around the reads, and can be partly modeled \citep[e.g.,]{Li10,roberts2011improving,wu2011using}.
However, accurate estimation of the rates is very difficult, especially
for paired-end data, and this inaccuracy brings extra variation that
needs to be included in the model. Therefore, we propose to use the
following negative binomial model, 
\begin{equation}
n_{j}|\theta\sim\mbox{NB}\left(\sum_{i=1}^{I}\theta_{i}a_{ij},\phi\right),\label{eq:nb_regression}
\end{equation}
 where NB is short for ``negative binomial'', the first term in
parentheses is the mean of the distribution, and $\phi$ is the dispersion
parameter so that $\mbox{var}(n_{j})=\mbox{mean}(n_{j})+\phi\cdot[\mbox{mean}(n_{j})]^{2}$.
Actually, since the estimation of rates are often difficult and cumbersome,
people tend to skip this step and use 1 or 0 for $a_{ij}$'s. In this
case, the counts are very heavily over-dispersed, and using a negative
binomial regression instead of Poisson is pressing.

This negative binomial regression model also helps take into account
of other biases and variations such as uncertainties in isoform annotations.
De novo assembly of transcriptomes often results in many isoforms
that have very low expression levels. Including all these transcriptomes
greatly increase the computational load and can cause non-identifiability
problems, and excluding them brings extra variations in the model.
De novo assembly as well as reference transcriptome can also have
mis-specified boundaries of some exons, which also brings extra variations
that need to be handled by the model. Using a negative binomial model
helps incorporating these biases/extra variations.

Recent years, negative binomial based generalized linear models (GLMs)
are extensively used for modeling RNA-Seq count data in the literatures
of identification of differentially expressed genes. Many state of
art methods, such as edgeR \citep{edgeR}, DESeq \citep{DESeq}, ShrinkBayes
\citep{ShrinkBayes}, and baySeq \citep{baySeq}, have been proposed
to estimate the coefficients and the dispersion parameters, and they
have had great success. These methods all use GLMs with log link,
while both Model \ref{eq:poisson_regression} and \ref{eq:nb_regression}
use identity link, which is required by the nature of isoform expression:
the expression of a gene or a part of a gene is the sum, not the product,
of isoforms. Unlike log link, identity link requires additional constraints
on the coefficients to make the mean of the Poisson distribution or
negative binomial distribution nonnegative, and thus brings difficulties
in estimating the coefficients.

\subsection{A robust quasi-likelihood estimator}

Model \ref{eq:nb_regression} is a negative-binomial regression model
with identity link and constraints $\theta_{i}\ge0$, $i=1,\ldots,I$.
We will discuss the estimation of the dispersion parameter $\phi$
in section \ref{sub:est_dispersion}. Here we assume $\phi$ is known
and the only parameters needs to be optimized is $\theta$. This optimization
is often done by maximizing the log-likelihood, which, however, gives
estimate that is very sensitive to outliers \citep[e.g.,]{pregibon1982resistant,stefanski1986optimally,kunsch1989conditionally,morgenthaler1992least,ruckstuhl2001robust}.
Outliers are generated by various reasons and are often common in
sequencing data \citep{ac2013reproducibility,SAM2}. It is worth noting
here the difference between ``biases/extra variations'' and ``outliers''.
The former are systematic uncertainties that affect a significant
proportion of counts (for example, inaccuracy in estimating $a_{ij}$
affects every $n_{j}$), and the latter are scattered and unpredictable
``errors'' that often affect only a small proportion of counts.
The dispersion parameter in negative binomial distribution is able
to efficiently taken into account the former but not the latter.

An efficient way to deal with outliers is to use robust estimators.
The theories of robust estimation is very well studied and widely
applied for ordinarily least squares, but less for generalized linear
models \citep[e.g.,]{hampel2011robust,huber2009robust,maronna2006robust}.
In a recent publication, \citet{edgeRrobust}, a robustified version
of the adjusted profile likelihood is proposed for negative-binomial-based
GLMs for identification of differentially expressed genes, but this
solution does not apply to identity link. We propose an approach based
on \citet{cantoni2001robust}, where the authors proposed a set of
M-estimators of Mallow's type that work on a large group of generalized
linear models, especially on binomial models and Poisson models. When
used on our negative binomial model, the estimator is given by the
solution to the following set of $I$ equations: 
\begin{equation}
\sum_{j=1}^{J}\nu(n_{j},\mu_{j})w_{j}a_{ij}-\sum_{j=1}^{J}\mathbb{E}[\nu(n_{j},\mu_{j})]w_{j}a_{ij}=0,\mbox{ for }i=1,\ldots,I.\label{eq:qusi_likelihood}
\end{equation}
 In the equations, 
\begin{itemize}
\item $\mu_{j}=\sum_{i=1}^{I}\theta_{i}a_{ij}$ is the expectation of $n_{j}$. 
\item $w_{j}$ is a pre-specified weight for the $j$'th row of matrix $A$.
In the literature of robust regression, $w_{j}$ is often set to be
a robust version of the Mahalanobis distances from the overall mean
of the rows of matrix $A$, or set to be all 1's. In this work, we
use the latter choice for simplicity. 
\item $\nu(n_{j},\mu_{j})=\frac{1}{\sqrt{V_{j}}}\cdot h\left(\frac{n_{j}-\mu_{j}}{\sqrt{V_{j}}}\right)$,
where $V_{j}=\mu_{j}+\phi\mu_{j}^{2}$ is the variance of $n_{j}$,
and $h$ is the first derivative of the Huber loss function, 
\[
h\left(\frac{n_{j}-\mu_{j}}{\sqrt{V_{j}}}\right)=\begin{cases}
\frac{n_{j}-\mu_{j}}{\sqrt{V_{j}}}, & \mbox{ if }\left|\frac{n_{j}-\mu_{j}}{\sqrt{V_{j}}}\right|\le c;\\
c\cdot\mbox{sign}(n_{j}-\mu_{j}), & \mbox{ otherwise}.
\end{cases}
\]
 Here $c$ is a positive constant, and $c=2.5$ is usually a reasonable
value \citep{rousseeuw2005robust}. 
\item $\mathbb{E}[\nu(n_{j},\mu_{j})]$ is the expectation of $\nu(n_{j},\mu_{j})$.
This term ensures the Fisher consistency of the estimator. Cantoni
and Ronchetti have shown that $\mathbb{E}[\nu(n_{j},\mu_{j})]$ has
a closed form in the case of binomial models and Poisson models \citep{cantoni2001robust}.
We have further shown that the following closed form also exists for
negative binomial distributions (See Appendix for details): $\mathbb{E}\left[\nu(n_{j},\mu_{j})\right]=\frac{c}{\sqrt{V_{j}}}(\Pr(Y_{j}\ge k_{j2}+1)-\Pr(Y_{j}\le k_{j1}))+\frac{\mu_{j}}{V_{j}}[\Pr(k_{j1}\le\tilde{Y}_{j}\le k_{j2}-1)-\Pr(k_{j1}+1\le Y_{j}\le k_{j2})],$
where $k_{j1}=\left\lfloor \mu_{j}-c\sqrt{V_{j}}\right\rfloor $,
$k_{j2}=\left\lfloor \mu_{j}+c\sqrt{V_{j}}\right\rfloor $, $Y_{j}\sim\mbox{NB}(\mu_{j},\phi)$,
and $\tilde{Y_{j}}\sim\mbox{NB}((1+\phi)\mu_{j},\frac{\phi}{\phi+1}).$
Here $\left\lfloor \cdot\right\rfloor $ means ``floor'', the largest
integer no greater than $\cdot$. 
\end{itemize}

\subsection{An algorithm to solve the quasi-likelihood equation}

Usually, the solution to GLMs are obtained by using the iterative
re-weighted least squares algorithm (IRLS). IRLS often works for log
link, but often fails for identity link with boundary constraints.
For Poisson GLMs with identity link (Model \ref{eq:poisson_regression}),
people have proposed to view the source of each reads as latent variables
and then an EM algorithm can be applied to find the maximum likelihood
estimation efficiently. However, such a simple EM algorithm is not
available for negative binomial distributions.

The solution of our model can be obtained by solving \ref{eq:qusi_likelihood},
a set of $I$ equations. However, with the constraints $\theta_{i}\ge0$,
there may not always be a solution satisfying the set of estimating
equations~\ref{eq:qusi_likelihood}. A better way that we have found
is to use the primitive function, if we view the left hand side of
\ref{eq:qusi_likelihood} as the first derivative to $\theta_{i}$,
as below:

\begin{equation}
Q=\sum_{j=1}^{J}\intop_{n_{j}}^{\mu_{j}}\left[\nu(n_{j},t)-\mathbb{E}[\nu(n_{j},t)]\right]\mbox{d}t\label{eq:prim_func}
\end{equation}
 Solving \ref{eq:qusi_likelihood} is equivalent to minimizing \ref{eq:prim_func}
with constraints $\theta_{i}\ge0$. This primitive function includes
$J$ one dimensional integrations that can be easily done numerically.
We have found providing both the primitive function \ref{eq:prim_func}
and the vector of first derivatives (the left hand sides of equations
\ref{eq:qusi_likelihood}), R function \texttt{optim} can find the
solution quickly and reliably by using the L-BFGS-B algorithm \citep{L_BFGS_B1, L_BFGS_B2}.
\texttt{optim} also requires a starting value of $\theta_{i}$, and
we have found using the regular maximum likelihood solution of Poisson
model, which can be easily achieved using the EM algorithm, works
nicely. We have successfully tested our optimization approach in thousands
of simulations with different parameter settings and random seeds,
as well as in a real RNA-Seq dataset with thousands of genes of different
structures and expression levels.

\subsection{Estimation of the dispersion parameter\label{sub:est_dispersion}}

Recent years, the estimation of the dispersion parameter for negative
binomial distribution has been studied extensively and state-of-art
methods have been proposed, especially for the problem of differential
expression identification \citep[e.g.,]{edgeR,DESeq}. They are, unfortunately,
usually not robust to outliers. Robust estimation of dispersion in
negative binomial regression models is generally a very difficult
problem, and we do not attempt to give a general solution. Instead,
we assume that the dispersion parameter $\phi$ is the same for all
genes, and we propose a method that works for genes with a unique
isoform. We then use the estimate of $\theta$ given by one-isoform
genes for all genes.

For a gene with only one isoform, our model becomes 
\[
n_{j}|\theta\sim\mbox{NB}\left(a_{j}\theta,\phi\right).
\]

We first try to get a robust estimate of $\theta$ regardless of the
value of $\phi$. Let $m_{j}=\frac{n_{j}}{a_{j}}$, then $\mathbb{E}\left(m_{j}\right)=\theta$,
which is the same for all $j$'s. We sort $m_{1},\ldots,m_{J}$ from
smallest to largest, then outliers, if exist, are likely to appear
on the two ends. To exclude them, we let $S_{\alpha}$ be the set
of $j$'s that $m_{j}$ is between the $\alpha$'th quantile and $(1-\alpha)$'th
quantile of $m_{1},\ldots,m_{J}$, with $\alpha$ being a pre-specified
constant $\in[0,0.25]$. Then we estimate $\theta$ by $\hat{\theta}=\sum_{j\in S_{\alpha}}n_{j}/\sum_{j\in S_{\alpha}}a_{j}$.
Simulations have shown that $\hat{\theta}$ is a very robust estimation
of $\theta$ given that the proportion of outliers does not exceed
$\alpha$.

Given $\hat{\theta}$, we use a moment estimator to estimate $\phi$.
Since $\mathbb{E}n_{j}=a_{j}\theta$ and $V(n_{j})=a_{j}\theta+\phi a_{j}^{2}\theta^{2}$,
we can estimate $\phi$ by $[\sum_{j=1}^{J}(n_{j}-a_{j}\hat{\theta})^{2}-\sum_{j=1}^{J}a_{j}\hat{\theta}]/\sum_{j=1}^{J}a_{j}^{2}\hat{\theta}^{2}$.
To robustify this estimator, we let $\beta$ be a pre-specified constant,
and $S_{\beta}$ be the set of $j$'s that $m_{j}$ is between the
$\alpha$'th quantile and $(1-\alpha)$'th quantile of $m_{1},\ldots,m_{J}$.
Then

\[
\hat{\phi}=\frac{\sum_{j\in S_{\beta}}(n_{j}-a_{j}\hat{\theta})^{2}-\sum_{j\in S_{\beta}}a_{j}\hat{\theta}}{\sum_{j\in S_{\beta}}a_{j}^{2}\hat{\theta}^{2}}.
\]
 This estimator turns out to underestimate $\theta$ as $S_{\beta}$
excludes $m_{j}$'s that are most diverse (even when there are actually
no outliers). To eliminate this bias, we calculate $\hat{\theta}$
under a series of $\beta$ that is no larger than $\alpha$, and then
fit a natural cubic spline on the relationship between $\hat{\theta}$
and $\beta$, and predict the value of $\hat{\theta}$ at the point
$\beta=0$. Simulations (Section \ref{subsec:simudisp}) have shown
that the resulted $\hat{\theta}$ robustly estimates $\theta$ with
relatively small bias.

In practice, we use the above method to estimate $\hat{\theta}$ for
every gene with only one isoform, and then use their average as the
estimated $\theta$ that will be used for all genes.

If one needs to use different dispersion parameters for different
genes, our methods works in one case: when the $A$ matrix composes
with 0's and 1's. As we have discussed in Section \ref{sub:nb_model},
this is the case when the users assign the same rate to all read types.
In this case, many columns of the $A$ matrix will be the same, and
we can use our method for $n_{j}$'s whose corresponding columns in
$A$ matrix are the same to estimate $\phi$, as these $n_{j}$'s
have the same mean.

In the general case that $a_{ij}$ are different from each other,
our method only works for gene with one isoform, but $\hat{\phi}$
estimated by them may be generalized to other genes if further assumptions
are made. For example, if one assume that $\phi$ is a smooth function
of gene expression, which is a common assumption in the literature
of differential expression identification \citep{DESeq}, then we
can estimate this smooth function using genes with one isoform. We
leave these to future research.

\section{Simulation results}

\subsection{Simulating data}

In this section, we assess the performance of our method and compare
it with other methods on simulation data with different gene structures,
sequencing depths, and levels of overdispersion. Data are simulated
according to \ref{eq:nb_regression} with different values of dispersion
$\phi$, isoform expression $\theta=(\theta_{1},\ldots,\theta_{I})$,
and sampling rate matrix $A=\{a_{ij}\}$. Three values of $\phi$
(0, 0.4, and 1) are used to represent no dispersion, moderate dispersion,
and strong dispersion. $A$ and $\theta$ are simulated using the
following four schemes:
\begin{enumerate}
\item \textit{genes with only one isoform}. We let $\theta=1$ and $A$=$bA^{\prime}$.
Here $A^{\prime}=(A_{1},\ldots,A_{50})$, where $A_{1},\ldots,A_{50}$
are independently generated from $\mbox{Uniform}(0.1,2)$. $b$ is
a constant equal to 10, 100 or 1000, corresponding to genes with small
read counts, moderate read counts, and large read counts. Note that
the read counts depends on the gene expression level and the sequencing
depth. After $(n_{1},\ldots,n_{J})$ are generated according to \ref{eq:nb_regression},
outliers are added: we let $n_{1}$ be 20 times of its expectation,
representing a very large value and we call it ``outlier to the right'',
or 0, representing a very small value and we call it ``outlier to
the left''. 
\item \textit{genes with two isoforms and both isoforms are expressed}.
We let $\theta=(\theta_{1},\theta_{2})=(0.8,0.2)$ and $A$=$bA^{\prime}$.
Here $A^{\prime}$ is a $2\times50$ matrix with all elements in the
first row and the first 25 elements in the second row independently
generated from $\mbox{Uniform}(0.1,2)$, and the last 25 elements
in the second row being 0. Again, $b$ is a constant equal to 10,
100 or 1000. We let $n_{1}$ be 20 times of its expectation or 0 to
represent outliers. 
\item \textit{genes with two isoforms and only one isoform is expressed}.
We use Scheme 2 to simulate data except letting $\theta=(\theta_{1},\theta_{2})=(1,0)$.
This represents the case when the solution is on the boundary of the
feasible region of our optimization problem \ref{eq:prim_func}. 
\item \textit{genes with five isoforms}. We generate $\vartheta_{1},\ldots,\vartheta_{5}$
from $\mbox{Uniform}(0,1)$ independently, and then set one of them
equals 0. Then we let $\theta_{i}=\vartheta_{i}/\sum_{k=1}^{5}\vartheta_{k}$,
and $\theta=(\theta_{1},\ldots,\theta_{5})$. $A$=$bA^{\prime}$,
where $b$ equals 10, 100, or 1000. $A^{\prime}$ is a $5\times100$
matrix, with its elements independently generated from $\mbox{Uniform}(0.1,2)$
with a half chance, or equals 0 otherwise. To add outliers, we let
$n_{1}$ and $n_{2}$ be 20 times of their expectations or 0's. 
\end{enumerate}

\subsection{Comparison of performance of different methods}

We ran our algorithm, as well as three other algorithms on the simulated
data: 
\begin{enumerate}
\item ``MLE'': proposed by \citet{jiang2009statistical}, this algorithm
gives the maximum likelihood estimate based on Poisson model \ref{eq:poisson_regression}.
It does not take outliers into account. 
\item ``Lasso1'': proposed by \citet{jiang2013penalized}, this algorithm
maximizes an L1-penalized log-likelihood function of Poisson model
\ref{eq:poisson_regression}. The L1-penalty identifies suspected
outliers and reduces their influence on the estimate. 
\item ``Lasso2'': also proposed by \citet{jiang2013penalized}, this algorithm
discards all counts that are detected as outliers by Lasso1 and then
calculates the maximum likelihood estimate of the other counts. Lasso2
fails in some simulations when the dispersion parameter is large,
as in this case all counts are detected as outliers by Lasso1 and
then discarded by Lasso2. We output the estimate of Lasso1 for Lasso2
in this case. 
\end{enumerate}
To measure the performance, we use $\mbox{RMSE}=\sqrt{\sum_{i=1}^{I}(\hat{\theta}_{i}-\theta_{i})^{2}/I}$,
where $\theta_{i}$ and $\hat{\theta}_{i}$ are the true and estimated
value of the expression of the $i$th isoform. Since in our simulation
we always let $\sum_{i=1}^{I}\theta_{i}=1$, this RMSE can be viewed
as the root mean squared error relative to the total expression of
all isoforms.

Tables \ref{tab:simu1} to \ref{tab:simu4} give the RMSE of all methods
under each of the four schemes. We did 100 simulations for each simulation
scheme, and report the mean and the standard error of the mean of
the 100 simulations. For short, we call our program ``R-QLE'', which
stands for robust quasi-likelihood estimate. The smallest RMSE in
each simulation scheme is marked as bold. The first impression is
that while MLE often gives the largest RMSE, there is no single method
that always gives the smallest RMSE. However, it is clear that our
method has the best overall performance.

Our method gives the smallest RMSE in 48 out of 72 (67\%) simulations.
Importantly, although in some simulations our method gives comparable
or a bit larger RMSE than Lasso1 or Lasso2, we haven't observed in
any of our simulations that our method gives an RMSE that is $>30\%$
larger than the best method. Only in 2 out of 72 (3\%) simluations
is our RMSE $>20\%$ larger than the best method, and 4 out of 72
(6\%) simulations is our RMSE $>10\%$ larger than the best method.
This means that the performance of our method is very reliable. Lasso1
and Lasso2 can give substantially larger RMSE than our method. In
17 and 5 out of 72 (24\% and 7\%) of simulations, Lasso1 gives RMSE
that is $>50\%$ and $>100\%$ larger than our method, respectively.
These two numbers are 27 and 12 out of 72 (38\% and 17\%) for Lasso2.
Especially, Lasso1 and Lasso2 tend to give much larger RMSE when the
dispersion parameter is median (0.4) or high (1), or $b$ is median
(100) to large (1000).

Additionally, we find that comparing the case of ``two isoforms,
both express'' and ``two isoforms, only one expresses'', the advantage
of our method is even larger in the latter case, indicating that our
method's reliability on the margin of feasible regions.

\subsection{Influence of the estimation of dispersion parameter \label{subsec:simudisp}}

In all the above simulations, we assume that the dispersion parameter
is known. For real data, as we have discussed in Section \ref{sub:est_dispersion},
we estimate the dispersion for each single-isoform gene, and use the
mean of estimated dispersion for all genes. We check the performance
of this strategy on simulation data. We simulate data according to
Scheme 1, and let 10\% of counts to be outliers. With a half chance,
these outliers are 20 times of the expected value, and 0 otherwise.
We simulate 100 genes as a group, and use the average of the estimated
dispersions as the final estimate of dispersion. Table \ref{tab:disp}
gives the mean and standard error of the mean based on 100 groups.
We see that the bias of the estimation is acceptably small.

\begin{center}
\begin{table}[H]
\caption{RMSE on simulation data: One isoform}

\label{tab:simu1} %
\begin{tabular}{llcccccccc}
 &  & \multicolumn{4}{c|}{outliers to the left} & \multicolumn{4}{c}{outliers to the right}\tabularnewline
 &  & R-QLE  & MLE  & Lasso1  & \multicolumn{1}{l}{Lasso2} & R-QLE  & MLE  & Lasso1  & Lasso2 \tabularnewline
\hline 
$b=10$  & $\phi=0$  & \textbf{.0373} & .0393 & .0383 & .0476 & \textbf{.0349} & .3909 & .0391 & .0351\tabularnewline
 &  & (.0026) & (.0028) & (.0026) & (.0035) & (.0029) & (.0199) & (.0032) & (.0027) \tabularnewline
\hline 
 & $\phi=0.4$  & \textbf{.0871} & .0935 & .1286 & .1667 & .0929 & .3920 & \textbf{.0891} & .1218\tabularnewline
 &  & (.0059) & (.0067) & (.0081) & (.0101) & (.0067) & (.0237) & (.0067) & (.0084) \tabularnewline
\hline 
 & $\phi=1$  & \textbf{.1200} & .1317 & .2671 & .3730 & \textbf{.1083} & .3052 & .1733 & .3049\tabularnewline
 &  & (.0084) & (.0092) & (.0110) & (.0134) & (.0096) & (.0222) & (.0092) & (.0121) \tabularnewline
\hline 
$b=100$  & $\phi=0$  & \textbf{.0133} & .0237 & .0135 & .0160 & \textbf{.0116} & .3568 & .0127 & .0124\tabularnewline
 &  & (.0011) & (.0017) & (.0011) & (.0011) & (.0010) & (.0184) & (.0010) & (.0010) \tabularnewline
\hline 
 & $\phi=0.4$  & \textbf{.0736} & .0840 & .1470 & .1675 & \textbf{.0780} & .3289 & .1068 & .1448\tabularnewline
 &  & (.0053) & (.0053) & (.0082) & (.0089) & (.0058) & (.0212) & (.0072) & (.0084) \tabularnewline
\hline 
 & $\phi=1$  & .1102 & \textbf{.1082} & .2892 & .3232 & \textbf{.1260} & .3567 & .2273 & .2808 \tabularnewline
 &  & (.0079) & (.0083) & (.0125) & (.0138) & (.0087) & (.0220) & (.0120) & (.0136) \tabularnewline
\hline 
$b=1000$  & $\phi=0$  & \textbf{.0034} & .0205 & .0036 & .0044 & .0033 & .3885 & .0040 & \textbf{.0033} \tabularnewline
 &  & (.0003) & (.0011) & (.0003) & (.0003) & (.0003) & (.0192) & (.0003) & (.0003) \tabularnewline
\hline 
 & $\phi=0.4$  & \textbf{.0679} & .0759 & .1348 & .1422 & \textbf{.0808} & .4050 & .1080 & .1156 \tabularnewline
 &  & (.0055) & (.0062) & (.0088) & (.0095) & (.0060) & (.0226) & (.0075) & (.0084) \tabularnewline
\hline 
 & $\phi=1$  & \textbf{.1163} & .1193 & .2984 & .3091 & \textbf{.1497} & .4180 & .2473 & .2619 \tabularnewline
 &  & (.0081) & (.0084) & (.0124) & (.0129) & (.0103) & (.0241) & (.0121) & (.0126) \tabularnewline
\end{tabular}
\end{table}

\par\end{center}

\begin{center}
\begin{table}[H]
\caption{RMSE on simulation data: Two isoforms, both expressed}

\label{tab:simu2} %
\begin{tabular}{llcccccccc}
 &  & \multicolumn{4}{c|}{outliers to the left} & \multicolumn{4}{c}{outliers to the right}\tabularnewline
 &  & R-QLE  & MLE  & Lasso1  & \multicolumn{1}{l}{Lasso2} & R-QLE  & MLE  & Lasso1  & Lasso2 \tabularnewline
\hline 
$b=10$  & $\phi=0$  & \textbf{.0481} & .0488 & .0490 & .0575 & .0529 & .4628 & .0605 & \textbf{.0465}\tabularnewline
 &  & (.0028) & (.0028) & (.0030) & (.0036) & (.0034) & (.0201) & (.0038) & (.0027) \tabularnewline
\hline 
 & $\phi=0.4$  & \textbf{.0910} & .0959 & .1031 & .1309 & .1145 & .4410 & \textbf{.1051} & .1079\tabularnewline
 &  & (.0052) & (.0050) & (.0048) & (.0059) & (.0073) & (.0207) & (.0059) & (.0052) \tabularnewline
\hline 
 & $\phi=1$  & .1453 & \textbf{.1421} & .1765 & .2289 & .1887 & .4870 & \textbf{.1561} & .1951\tabularnewline
 &  & (.0081) & (.0091) & (.0065) & (.0075) & (.0130) & (.0251) & (.0080) & (.0080) \tabularnewline
\hline 
$b=100$  & $\phi=0$  & .0169 & .0286 & \textbf{.0166} & .0179 & .0163 & .4295 & .0183 & \textbf{.0154}\tabularnewline
 &  & (.0008) & (.0014) & (.0009) & (.0009) & (.0009) & (.0184) & (.0011) & (.0008) \tabularnewline
\hline 
 & $\phi=0.4$  & \textbf{.0906} & .0951 & .1141 & .1294 & .1100 & .4415 & \textbf{.1026} & .1240\tabularnewline
 &  & (.0043) & (.0050) & (.0050) & (.0058) & (.0071) & (.0203) & (.0060) & (.0067) \tabularnewline
\hline 
 & $\phi=1$  & \textbf{.1346} & .1382 & .2036 & .2273 & .1808 & .4333 & \textbf{.1796} & .2148 \tabularnewline
 &  & (.0068) & (.0073) & (.0071) & (.0083) & (.0100) & (.0193) & (.0075) & (.0088) \tabularnewline
\hline 
$b=1000$  & $\phi=0$  & \textbf{.0055} & .0245 & .0057 & .0065 & .0051 & .4319 & .0058 & \textbf{.0049} \tabularnewline
 &  & (.0003) & (.0012) & (.0003) & (.0004) & (.0003) & (.0202) & (.0004) & (.0003) \tabularnewline
\hline 
 & $\phi=0.4$  & \textbf{.0880} & .0974 & .1157 & .1211 & \textbf{.1085} & .4838 & .1114 & .1254 \tabularnewline
 &  & (.0044) & (.0049) & (.0051) & (.0053) & (.0065) & (.0225) & (.0059) & (.0064) \tabularnewline
\hline 
 & $\phi=1$  & .1327 & \textbf{.1323} & .2053 & .2195 & \textbf{.1626} & .4592 & .1857 & .2026 \tabularnewline
 &  & (.0073) & (.0080) & (.0073) & (.0087) & (.0110) & (.0184) & (.0079) & (.0085) \tabularnewline
\end{tabular}
\end{table}

\par\end{center}

\begin{center}
\begin{table}[H]
\caption{RMSE on simulation data: Two isoforms, only one expressed}

\label{tab:simu3} %
\begin{tabular}{llcccccccc}
 &  & \multicolumn{4}{c|}{outliers to the left} & \multicolumn{4}{c}{outliers to the right}\tabularnewline
 &  & R-QLE  & MLE  & Lasso1  & \multicolumn{1}{l}{Lasso2} & R-QLE  & MLE  & Lasso1  & Lasso2 \tabularnewline
\hline 
$b=10$  & $\theta=0$  & \textbf{.0342} & .0361 & .0351 & .0453 & .0402 & .4163 & .0468 & \textbf{.0354}\tabularnewline
 &  & (.0029) & (.0030) & (.0030) & (.0036) & (.0030) & (.0175) & (.0033) & (.0030) \tabularnewline
\hline 
 & $\theta=0.4$  & \textbf{.0754} & .0847 & .1095 & .1485 & \textbf{.0993} & .4319 & .1036 & .1270\tabularnewline
 &  & (.0056) & (.0058) & (.0070) & (.0079) & (.0065) & (.0209) & (.0067) & (.0077) \tabularnewline
\hline 
 & $\theta=1$  & \textbf{.1245} & .1368 & .2092 & .2718 & \textbf{.1475} & .4164 & .1619 & .2395\tabularnewline
 &  & (.0088) & (.0105) & (.0104) & (.0117) & (.0122) & (.0188) & (.0113) & (.0110) \tabularnewline
\hline 
$b=100$  & $\theta=0$  & \textbf{.0101} & .0177 & .0102 & .0121 & .0114 & .4037 & .0133 & \textbf{.0104}\tabularnewline
 &  & (.0006) & (.0011) & (.0006) & (.0008) & (.0007) & (.0186) & (.0009) & (.0007) \tabularnewline
\hline 
 & $\theta=0.4$  & \textbf{.0699} & .0751 & .1259 & .1402 & \textbf{.0842} & .3851 & .1133 & .1332\tabularnewline
 &  & (.0043) & (.0057) & (.0071) & (.0077) & (.0060) & (.0179) & (.0067) & (.0079) \tabularnewline
\hline 
 & $\theta=1$  & \textbf{.1215} & .1322 & .2499 & .2731 & \textbf{.1524} & .4701 & .2247 & .2697 \tabularnewline
 &  & (.0081) & (.0092) & (.0102) & (.0108) & (.0114) & (.0274) & (.0103) & (.0115) \tabularnewline
\hline 
$b=1000$  & $\theta=0$  & \textbf{.0038} & .0162 & .0039 & .0047 & .0040 & .4094 & .0048 & \textbf{.0037} \tabularnewline
 &  & (.0003) & (.0009) & (.0003) & (.0003) & (.0003) & (.0191) & (.0004) & (.0003) \tabularnewline
\hline 
 & $\theta=0.4$  & \textbf{.0610} & .0745 & .1179 & .1283 & \textbf{.0742} & .4391 & .1009 & .1139 \tabularnewline
 &  & (.0045) & (.0055) & (.0061) & (.0066) & (.0053) & (.0186) & (.0063) & (.0070) \tabularnewline
\hline 
 & $\theta=1$  & \textbf{.0964} & .1134 & .2430 & .2522 & \textbf{.1387} & .4184 & .2245 & .2403 \tabularnewline
 &  & (.0071) & (.0072) & (.0093) & (.0100) & (.0109) & (.0201) & (.0098) & (.0111) \tabularnewline
\end{tabular}
\end{table}

\par\end{center}

\begin{center}
\begin{table}[H]
\caption{RMSE on simulation data: Five isoforms}

\label{tab:simu4} %
\begin{tabular}{llcccccccc}
 &  & \multicolumn{4}{c|}{outliers to the left} & \multicolumn{4}{c}{outliers to the right}\tabularnewline
 &  & R-QLE  & MLE  & Lasso1  & \multicolumn{1}{l}{Lasso2} & R-QLE  & MLE  & Lasso1  & Lasso2 \tabularnewline
\hline 
$b=10$  & $\phi=0$  & .0288 & \textbf{.0287} & .0289 & .0330 & .0305 & .1646 & .0360 & \textbf{.0279}\tabularnewline
 &  & (.0013) & (.0013) & (.0014) & (.0015) & (.0012) & (.0066) & (.0014) & (.0013) \tabularnewline
\hline 
 & $\phi=0.4$  & \textbf{.0443} & .0461 & .0488 & .0633 & .0535 & .1766 & .0506 & \textbf{.0487}\tabularnewline
 &  & (.0021) & (.0020) & (.0022) & (.0023) & (.0023) & (.0076) & (.0020) & (.0022) \tabularnewline
\hline 
 & $\phi=1$  & \textbf{.0676} & .0701 & .0728 & .1016 & .0829 & .1959 & \textbf{.0638} & .0811\tabularnewline
 &  & (.0029) & (.0028) & (.0022) & (.0028) & (.0041) & (.0071) & (.0027) & (.0025) \tabularnewline
\hline 
$b=100$  & $\phi=0$  & \textbf{.0086} & .0124 & .0087 & .0094 & .0087 & .1588 & .0112 & \textbf{.0080}\tabularnewline
 &  & (.0003) & (.0005) & (.0003) & (.0004) & (.0003) & (.0064) & (.0004) & (.0003) \tabularnewline
\hline 
 & $\phi=0.4$  & \textbf{.0400} & .0464 & .0506 & .0561 & \textbf{.0442} & .1642 & .0442 & .0531\tabularnewline
 &  & (.0016) & (.0019) & (.0018) & (.0019) & (.0020) & (.0063) & (.0017) & (.0018) \tabularnewline
\hline 
 & $\phi=1$  & \textbf{.0597} & .0678 & .0865 & .0967 & .0796 & .1979 & \textbf{.0743} & .0928 \tabularnewline
 &  & (.0027) & (.0028) & (.0029) & (.0031) & (.0036) & (.0075) & (.0029) & (.0033) \tabularnewline
\hline 
$b=1000$  & $\phi=0$  & .0028 & .0101 & \textbf{.0028} & .0031 & .0029 & .1617 & .0038 & \textbf{.0027} \tabularnewline
 &  & (.0001) & (.0004) & (.0001) & (.0001) & (.0001) & (.0066) & (.0002) & (.0001) \tabularnewline
\hline 
 & $\phi=0.4$  & \textbf{.0403} & .0453 & .0545 & .0581 & \textbf{.0442} & .1709 & .0492 & .0545 \tabularnewline
 &  & (.0016) & (.0019) & (.0019) & (.0019) & (.0019) & (.0067) & (.0019) & (.0020) \tabularnewline
\hline 
 & $\phi=1$  & \textbf{.0532} & .0606 & .0907 & .0947 & \textbf{.0638} & .1731 & .0802 & .0880 \tabularnewline
 &  & (.0026) & (.0024) & (.0027) & (.0028) & (.0032) & (.0060) & (.0026) & (.0028) \tabularnewline
\end{tabular}
\end{table}

\par\end{center}

\begin{center}
\begin{table}[h]
\caption{Estimation of dispersion parameters}

\label{tab:disp} %
\begin{tabular}{llll}
 & $b=10$  & $b=100$  & $b=1000$ \tabularnewline
$\phi=0.2$  & 0.2048 (0.0117)  & 0.1714 (0.0059)  & 0.1678 (0.0059) \tabularnewline
$\phi=0.4$  & 0.4070 (0.0164)  & 0.3482 (0.0131)  & 0.3427 (0.0132) \tabularnewline
$\phi=0.6$  & 0.6155 (0.0252)  & 0.5207 (0.0230)  & 0.5207 (0.0200) \tabularnewline
\end{tabular}
\end{table}

\par\end{center}

We assume that all genes have the same dispersion parameter. This
assumption can be strong for real data. So we study the performance
of our method when an inaccurate dispersion parameter is used. We
simulate data under Scheme 4 (five isoforms) using three different
dispersions 0, 0.4, and 1, and estimate the dispersion using an inaccurate
estimation of dispersion, 0.3. Table \ref{tab:simuf4} gives the mean
and the standard error of mean under 100 simulations.

Comparing Table \ref{tab:simuf4} with Table \ref{tab:simu4}, of
course the performance of MLE, Lasso1, and Lasso2 do not change, as
they do not use the dispersion. The RMSE of our method increases significantly
in the cases when the true dispersion is 0 and $b=100\mbox{ or }1000$.
This is easy to understand, as in this case, the true outliers will
not be regarded as outliers when one assumes the dispersion is 0.4,
a much larger value than the true value. Nevertheless, the RMSE is
still small comparing with the RMSEs under simulation data with larger
dispersions. When the data is simulated under $\phi=0.4$ or 1, the
RMSE of our method does not increase significantly, and it still outperforms
other methods in many cases. The comparisons under the other three
simulation schemes give similar conclusions.

\begin{center}
\begin{table}[h]
\caption{RMSE on simulation data (using $\phi=0.3$ for estimation): Five isoforms}

\label{tab:simuf4} %
\begin{tabular}{llcccccccc}
 &  & \multicolumn{4}{c|}{outliers to the left} & \multicolumn{4}{c}{outliers to the right}\tabularnewline
 &  & R-QLE  & MLE  & Lasso1  & \multicolumn{1}{l}{Lasso2} & R-QLE  & MLE  & Lasso1  & Lasso2 \tabularnewline
\hline 
$b=10$  & $\phi=0$  & .0296 & \textbf{.0287} & .0289 & .0330 & .0368 & .1646 & .0360 & \textbf{.0279}\tabularnewline
 &  & (.0014) & (.0013) & (.0014) & (.0015) & (.0015) & (.0066) & (.0014) & (.0013) \tabularnewline
\hline 
 & $\phi=0.4$  & \textbf{.0442} & .0461 & .0488 & .0633 & .0514 & .1766 & .0506 & \textbf{.0487}\tabularnewline
 &  & (.0021) & (.0020) & (.0022) & (.0023) & (.0022) & (.0076) & (.0020) & (.0022) \tabularnewline
\hline 
 & $\phi=1$  & \textbf{.0643} & .0701 & .0728 & .1016 & .0664 & .1959 & \textbf{.0638} & .0811\tabularnewline
 &  & (.0025) & (.0028) & (.0022) & (.0028) & (.0031) & (.0071) & (.0027) & (.0025) \tabularnewline
\hline 
$b=100$  & $\phi=0$  & .0125 & .0124 & \textbf{.0087} & .0094 & .0188 & .1588 & .0112 & \textbf{.0080}\tabularnewline
 &  & (.0005) & (.0005) & (.0003) & (.0004) & (.0006) & (.0064) & (.0004) & (.0003) \tabularnewline
\hline 
 & $\phi=0.4$  & \textbf{.0402} & .0464 & .0506 & .0561 & \textbf{.0427} & .1642 & .0442 & .0531\tabularnewline
 &  & (.0016) & (.0019) & (.0018) & (.0019) & (.0019) & (.0063) & (.0017) & (.0018) \tabularnewline
\hline 
 & $\phi=1$  & \textbf{.0639} & .0678 & .0865 & .0967 & \textbf{.0662} & .1979 & .0743 & .0928 \tabularnewline
 &  & (.0026) & (.0028) & (.0029) & (.0031) & (.0029) & (.0075) & (.0029) & (.0033) \tabularnewline
\hline 
$b=1000$  & $\phi=0$  & .0087 & .0101 & \textbf{.0028} & .0031 & .0158 & .1617 & .0038 & \textbf{.0027} \tabularnewline
 &  & (.0004) & (.0004) & (.0001) & (.0001) & (.0006) & (.0066) & (.0002) & (.0001) \tabularnewline
\hline 
 & $\phi=0.4$  & \textbf{.0407} & .0453 & .0545 & .0581 & \textbf{.0432} & .1709 & .0492 & .0545 \tabularnewline
 &  & (.0016) & (.0019) & (.0019) & (.0019) & (.0019) & (.0067) & (.0019) & (.0020) \tabularnewline
\hline 
 & $\phi=1$  & \textbf{.0586} & .0606 & .0907 & .0947 & \textbf{.0549} & .1731 & .0802 & .0880 \tabularnewline
 &  & (.0023) & (.0024) & (.0027) & (.0028) & (.0027) & (.0060) & (.0026) & (.0028) \tabularnewline
\end{tabular}
\end{table}

\par\end{center}

\section{Real data analysis}

For real data analysis, we use RNA-Seq data from the H1 human embryonic
stem cell line generated by the Cold Spring Harbor Laboratory in the
ENCODE project \citep{ENCODE}. A total of 78 million single-end reads
of 75 bp mapped to the RefSeq human annotation database \citep{Pruitt2009}
are used in the analysis. We apply the same four algorithms as in
the simulated data analysis: R-QLE, MLE, Lasso1 and Lasso2.

For each of the $20,297$ annotated genes, based on its annotated
isoforms, we count the number of reads mapped to each exons or junctions,
which we define as read types. We then apply our algorithm for estimating
the dispersion parameter to genes with only one isoform, at least
20 read types and a median read count across all read types of at
least 10. A total of $1,751$ genes are used for the estimation. The
mean of estimated dispersion parameters is $0.304$, which is used
as the dispersion parameter in R-QLE for later analysis.

We estimate the isoform expression values using the four algorithms
for a total of $13,272$ genes having at least $100$ mapped reads.
We then estimate the gene expression value as the sum of expression
values of all its isoforms and use the gene expression value as the
basis for comparisons across different methods. Overall, all the four
algorithms give quite concordant results. For example, the Spearman
correlation coefficients between the estimates given by R-QLE and
three other algorithms (MLE, Lasso1 and Lasso2) are $0.922$, $0.981$
and $0.938$, respectively. Only $292$ genes have a change larger
than $2$ folds between their estimates given by R-QLE and MLE.

To compare the robustness of the four algorithms, for each gene, we
remove a quarter of the observed data, by removing the first quarter
of elements in $N$ and the first quarter of columns in $A$ correspondingly,
and re-estimate isoform expression values using the four algorithms.
The rationale is that a more robust method should be less affected
when part of the observations are removed. The Spearman correlation
coefficients between the estimates based on the complete data and
partial data, using the four algorithms, are $0.978$ (R-QLE) $>0.965$
(Lasso1) $>0.951$ (MLE) $>0.941$ (Lasso2). We can see that R-QLE
clearly outperforms all three other methods. The difference becomes
even larger when we only focus on the $292$ genes with a change larger
than $2$ folds: $0.978$ (R-QLE) $>0.955$ (Lasso1) $>0.924$ (MLE)
$>0.923$ (Lasso2).

\section{Conclusion}

The most commonly considered GLMs for counts data are Poisson distribution
based and with log link, which is numerically easy to deal with. However,
the nature of isoform expression based on RNA-Seq data requires a
GLM based on negative binomial distribution, with identity link, and
robust to outliers. There has not been any successful example of such
GLM models, and one reason can be the difficulty in optimizing the
coefficients. We have identified a numeric algorithm that appeared
to be both efficient and reliable. Simulation results show that the
estimate of isoform expression from our method is more accurate and
reliable than existing methods, and the reliability of our method
is also shown in real data.

\section{Appendix section}

\label{app}

In this appendix, we give the closed form for $\mathbb{E}[\nu(n_{j},\mu_{j})]$.

We would like to find the expectation of 
\begin{eqnarray*}
\nu(n_{j},\mu_{j}) & = & \begin{cases}
\frac{n_{j}-\mu_{j}}{V_{j}}, & \mbox{ if }\left|\frac{n_{j}-\mu_{j}}{\sqrt{V_{j}}}\right|\le c;\\
\frac{c}{\sqrt{V_{j}}}\mbox{sign}(n_{j}-\mu_{j}), & \mbox{ otherwise}.
\end{cases}
\end{eqnarray*}
 Here $n_{j}\sim\mbox{NB}(\mu_{j},\phi)$. To simplify the notation,
we write $n_{j}$ as $n$, $\mu_{j}$ as $\mu$, and $V_{j}$ as $V$.
Let $k_{1}=\left\lfloor \mu-c\sqrt{V}\right\rfloor $ and $k_{2}=\left\lfloor \mu+c\sqrt{V}\right\rfloor $,
and $Y$ be a random variable that follows $\mbox{NB}(\mu,\phi)$
distribution, then 
\begin{eqnarray*}
\mathbb{E}\left[\nu(n,\mu)\right] & = & \sum_{s=k_{1}+1}^{k_{2}}\frac{s-\mu}{V}\Pr(Y=s)-\frac{c}{\sqrt{V}}\sum_{s=0}^{k_{1}}\Pr(Y=s)+\frac{c}{\sqrt{V}}\sum_{s=k_{2}+1}^{+\infty}\Pr(Y=s)\\
 & = & \frac{1}{V}\sum_{s=k_{1}+1}^{k_{2}}s\Pr(Y=s)-\frac{\mu}{V}\Pr(k_{1}+1\le Y\le k_{2})\\
 &  & +\frac{c}{\sqrt{V}}(\Pr(Y\ge k_{2}+1)-\Pr(Y\le k_{1}))
\end{eqnarray*}
 We want to find a simple form for $M\triangleq\sum_{s=k_{1}+1}^{k_{2}}s\Pr(Y=s)$.
Plugging the probability density function of the negative binomial
distribution, we have

\begin{eqnarray*}
M & = & \frac{1}{\Gamma(\phi^{-1})}\left(\frac{\phi^{-1}}{\mu+\phi^{-1}}\right)^{\phi^{-1}}\sum_{s=k_{1}+1}^{k_{2}}s\frac{\Gamma(s+\phi^{-1})}{\Gamma(s+1)}\left(\frac{\mu}{\mu+\phi^{-1}}\right)^{s}\\
 & = & \frac{1}{\Gamma(\phi^{-1})}\left(\frac{\phi^{-1}}{\mu+\phi^{-1}}\right)^{\phi^{-1}}\sum_{s=k_{1}+1}^{k_{2}}\frac{\Gamma(s+\phi^{-1})}{\Gamma(s)}\left(\frac{\mu}{\mu+\phi^{-1}}\right)^{s}
\end{eqnarray*}

Let $\phi^{\prime-1}=1+\phi^{-1}$ and $\frac{\mu^{\prime}}{\mu^{\prime}+\phi^{\prime-1}}=\frac{\mu}{\mu+\phi^{-1}}$,
that is, $\phi^{\prime}=\frac{\phi}{\phi+1}$ and $\mu^{\prime}=(1+\phi)\mu$.
Then 
\begin{eqnarray*}
M & = & \frac{1}{\Gamma(\phi^{-1})}\left(\frac{\phi^{-1}}{\mu+\phi^{-1}}\right)^{\phi^{-1}}\sum_{s=k_{1}+1}^{k_{2}}\frac{\Gamma(s-1+\phi^{\prime-1})}{\Gamma(s-1+1)}\left(\frac{\mu^{\prime}}{\mu^{\prime}+\phi^{\prime-1}}\right)^{s-1+1}\\
 & = & \frac{1}{\Gamma(\phi^{-1})}\left(\frac{\phi^{-1}}{\mu+\phi^{-1}}\right)^{\phi^{-1}}\left(\frac{\mu^{\prime}}{\mu^{\prime}+\phi^{\prime-1}}\right)\sum_{s=k_{1}}^{k_{2}-1}\frac{\Gamma(s+\phi^{\prime-1})}{\Gamma(s+1)}\left(\frac{\mu^{\prime}}{\mu^{\prime}+\phi^{\prime-1}}\right)^{s}\\
 & = & \left[\frac{\Gamma(\phi^{\prime-1})}{\Gamma(\phi^{-1})}\left(\frac{\phi^{-1}}{\mu+\phi^{-1}}\right)^{\phi^{-1}}\left(\frac{\mu^{\prime}}{\mu^{\prime}+\phi^{\prime-1}}\right)\left(\frac{\phi^{\prime-1}}{\mu^{\prime}+\phi^{\prime-1}}\right)^{-\phi^{\prime-1}}\right]\\
 &  & \cdot\left[\sum_{s=k_{1}}^{k_{2}-1}\frac{\Gamma(s+\phi^{\prime-1})}{\Gamma(\phi^{\prime-1})\Gamma(s+1)}\left(\frac{\phi^{\prime-1}}{\mu^{\prime}+\phi^{\prime-1}}\right)^{\phi^{\prime-1}}\left(\frac{\mu^{\prime}}{\mu^{\prime}+\phi^{\prime-1}}\right)^{s}\right]
\end{eqnarray*}
 The elements in the sum of the second pair of brackets is the probability
density function of $\mbox{NB}((1+\phi)\mu,\frac{\phi}{\phi+1})$.
Thus, the sum in the second pair of brackets equals $\Pr(k_{1}\le\tilde{Y}\le k_{2}-1)$,
where $\tilde{Y}\sim\mbox{NB}((1+\phi)\mu,\frac{\phi}{\phi+1})$.
It is also easy to show that the part in the first pair of brackets
can be simplified to $\mu$. Therefore, $M=\mu\Pr(k_{1}\le\tilde{Y}\le k_{2}-1)$,
and 
\begin{eqnarray*}
\mathbb{E}\left[\nu(n,\mu)\right] & = & \frac{\mu}{V}\left[\Pr(k_{1}\le\tilde{Y}\le k_{2}-1)-\Pr(k_{1}+1\le Y\le k_{2})\right]\\
 &  & +\frac{c}{\sqrt{V}}(\Pr(Y\ge k_{2}+1)-\Pr(Y\le k_{1})).
\end{eqnarray*}

\section*{Acknowledgements}

J.L. is supported by University of Notre Dame (startup grant). H.J.
is supported by University of Michigan (startup grant).

 \bibliographystyle{plainnat}

\begin{thebibliography}{32}
	\providecommand{\natexlab}[1]{#1}
	\providecommand{\url}[1]{\texttt{#1}}
	\expandafter\ifx\csname urlstyle\endcsname\relax
	\providecommand{\doi}[1]{doi: #1}\else
	\providecommand{\doi}{doi: \begingroup \urlstyle{rm}\Url}\fi
	
	\bibitem[AC't~Hoen et~al.(2013)AC't~Hoen, Friedl{\"a}nder, Alml{\"o}f, Sammeth,
	Pulyakhina, Anvar, Laros, Buermans, Karlberg, Br{\"a}nnvall,
	et~al.]{ac2013reproducibility}
	Peter AC't~Hoen, Marc~R Friedl{\"a}nder, Jonas Alml{\"o}f, Michael Sammeth,
	Irina Pulyakhina, Seyed~Yahya Anvar, Jeroen~FJ Laros, Henk~PJ Buermans, Olof
	Karlberg, Mathias Br{\"a}nnvall, et~al.
	\newblock Reproducibility of high-throughput mrna and small rna sequencing
	across laboratories.
	\newblock \emph{Nature biotechnology}, 2013.
	
	\bibitem[Anders and Huber(2010)]{DESeq}
	S.~Anders and W.~Huber.
	\newblock Differential expression analysis for sequence count data.
	\newblock \emph{Genome Biology}, 11:\penalty0 R106, 2010.
	
	\bibitem[Byrd et~al.(1995)Byrd, Lu, Nocedal, and Zhu]{L_BFGS_B1}
	Richard~H Byrd, Peihuang Lu, Jorge Nocedal, and Ciyou Zhu.
	\newblock A limited memory algorithm for bound constrained optimization.
	\newblock \emph{SIAM Journal on Scientific Computing}, 16\penalty0
	(5):\penalty0 1190--1208, 1995.
	
	\bibitem[Cantoni and Ronchetti(2001)]{cantoni2001robust}
	Eva Cantoni and Elvezio Ronchetti.
	\newblock Robust inference for generalized linear models.
	\newblock \emph{Journal of the American Statistical Association}, 96\penalty0
	(455), 2001.
	
	\bibitem[Consortium et~al.(2004)]{ENCODE}
	ENCODE~Project Consortium et~al.
	\newblock The encode (encyclopedia of dna elements) project.
	\newblock \emph{Science}, 306\penalty0 (5696):\penalty0 636--640, 2004.
	
	\bibitem[Hampel et~al.(2011)Hampel, Ronchetti, Rousseeuw, and
	Stahel]{hampel2011robust}
	Frank~R Hampel, Elvezio~M Ronchetti, Peter~J Rousseeuw, and Werner~A Stahel.
	\newblock \emph{Robust statistics: the approach based on influence functions},
	volume 114.
	\newblock John Wiley \& Sons, 2011.
	
	\bibitem[Hansen et~al.(2010)Hansen, Brenner, and Dudoit]{Hansen10}
	K.~D. Hansen, S.~E. Brenner, and S.~Dudoit.
	\newblock Biases in illumina transcriptome sequencing caused by random hexamer
	priming.
	\newblock \emph{Nucleic Acids Research}, 38:\penalty0 e131, 2010.
	
	\bibitem[Hardcastle and Kelly(2010)]{baySeq}
	T.~J. Hardcastle and K.~A. Kelly.
	\newblock bayseq: Empirical bayesian methods for identifying differential
	expression in sequence count data.
	\newblock \emph{BMC Bioinformatics}, 11:\penalty0 422, 2010.
	
	\bibitem[Huber and Ronchetti(2009)]{huber2009robust}
	Peter~J Huber and Elvezio~M Ronchetti.
	\newblock \emph{Robust Statistics}.
	\newblock Wiley, 2009.
	
	\bibitem[Jiang and Salzman(2013)]{jiang2013penalized}
	Hui Jiang and Julia Salzman.
	\newblock A penalized likelihood approach for robust estimation of isoform
	expression.
	\newblock \emph{arXiv preprint arXiv:1310.0379}, 2013.
	
	\bibitem[Jiang and Wong(2009)]{jiang2009statistical}
	Hui Jiang and Wing~Hung Wong.
	\newblock Statistical inferences for isoform expression in rna-seq.
	\newblock \emph{Bioinformatics}, 25\penalty0 (8):\penalty0 1026--1032, 2009.
	
	\bibitem[K{\"u}nsch et~al.(1989)K{\"u}nsch, Stefanski, and
	Carroll]{kunsch1989conditionally}
	Hans~R K{\"u}nsch, Leonard~A Stefanski, and Raymond~J Carroll.
	\newblock Conditionally unbiased bounded-influence estimation in general
	regression models, with applications to generalized linear models.
	\newblock \emph{Journal of the American Statistical Association}, 84\penalty0
	(406):\penalty0 460--466, 1989.
	
	\bibitem[Li and Dewey(2011)]{li2011rsem}
	Bo~Li and Colin~N Dewey.
	\newblock Rsem: accurate transcript quantification from rna-seq data with or
	without a reference genome.
	\newblock \emph{BMC bioinformatics}, 12\penalty0 (1):\penalty0 323, 2011.
	
	\bibitem[Li and Tibshirani(2011)]{SAM2}
	J.~Li and R.~Tibshirani.
	\newblock Finding consistent patterns: a nonparametric approach for identifying
	differential expression in rna-seq data.
	\newblock \emph{Statistical Methods in Medical Research}, 2011.
	\newblock To appear.
	
	\bibitem[Li et~al.(2010)Li, Jiang, and Wong]{Li10}
	J.~Li, H.~Jiang, and W.~H. Wong.
	\newblock Modeling non-uniformity in short-read rates in rna-seq data.
	\newblock \emph{Genome Biol}, 11\penalty0 (5):\penalty0 R50, 2010.
	
	\bibitem[L{\'o}pez-Bigas et~al.(2005)L{\'o}pez-Bigas, Audit, Ouzounis, Parra,
	and Guig{\'o}]{lopez2005splicing}
	N{\'u}ria L{\'o}pez-Bigas, Benjamin Audit, Christos Ouzounis, Gen{\'\i}s Parra,
	and Roderic Guig{\'o}.
	\newblock Are splicing mutations the most frequent cause of hereditary disease?
	\newblock \emph{FEBS letters}, 579\penalty0 (9):\penalty0 1900--1903, 2005.
	
	\bibitem[Maronna et~al.(2006)Maronna, Martin, and Yohai]{maronna2006robust}
	Ricardo Maronna, Douglas Martin, and Victor Yohai.
	\newblock \emph{Robust statistics}.
	\newblock John Wiley \& Sons, Chichester. ISBN, 2006.
	
	\bibitem[Morgenthaler(1992)]{morgenthaler1992least}
	Stephan Morgenthaler.
	\newblock Least-absolute-deviations fits for generalized linear models.
	\newblock \emph{Biometrika}, 79\penalty0 (4):\penalty0 747--754, 1992.
	
	\bibitem[Pachter(2011)]{pachter2011models}
	Lior Pachter.
	\newblock Models for transcript quantification from rna-seq.
	\newblock \emph{arXiv preprint arXiv:1104.3889}, 2011.
	
	\bibitem[Pregibon(1982)]{pregibon1982resistant}
	Daryl Pregibon.
	\newblock Resistant fits for some commonly used logistic models with medical
	applications.
	\newblock \emph{Biometrics}, pages 485--498, 1982.
	
	\bibitem[Pruitt et~al.(2009)Pruitt, Tatusova, Klimke, and Maglott]{Pruitt2009}
	Kim~D. Pruitt, Tatiana Tatusova, William Klimke, and Donna~R. Maglott.
	\newblock Ncbi reference sequences: current status, policy and new initiatives.
	\newblock \emph{Nucleic Acids Res}, 37\penalty0 (Database issue):\penalty0
	D32--D36, Jan 2009.
	
	\bibitem[Roberts et~al.(2011)Roberts, Trapnell, Donaghey, Rinn, and
	Pachter]{roberts2011improving}
	Adam Roberts, Cole Trapnell, Julie Donaghey, John~L Rinn, and Lior Pachter.
	\newblock Improving rna-seq expression estimates by correcting for fragment
	bias.
	\newblock \emph{Genome biology}, 12\penalty0 (3):\penalty0 R22, 2011.
	
	\bibitem[Robinson et~al.(2010)Robinson, McCarthy, and Smyth]{edgeR}
	M.~D. Robinson, D.~J. McCarthy, and G.~K. Smyth.
	\newblock edger: a bioconductor package for differential expression analysis of
	digital gene expression data.
	\newblock \emph{Bioinformatics}, 26\penalty0 (1):\penalty0 139--40, 2010.
	
	\bibitem[Rousseeuw and Leroy(2005)]{rousseeuw2005robust}
	Peter~J Rousseeuw and Annick~M Leroy.
	\newblock \emph{Robust regression and outlier detection}, volume 589.
	\newblock John Wiley \& Sons, 2005.
	
	\bibitem[Ruckstuhl and Welsh(2001)]{ruckstuhl2001robust}
	AF~Ruckstuhl and AH~Welsh.
	\newblock Robust fitting of the binomial model.
	\newblock \emph{Annals of statistics}, pages 1117--1136, 2001.
	
	\bibitem[Salzman et~al.(2011)Salzman, Jiang, and Wong]{salzman2011statistical}
	Julia Salzman, Hui Jiang, and Wing~Hung Wong.
	\newblock Statistical modeling of rna-seq data.
	\newblock \emph{Statistical science: a review journal of the Institute of
		Mathematical Statistics}, 26\penalty0 (1), 2011.
	
	\bibitem[Stefanski et~al.(1986)Stefanski, Carroll, and
	Ruppert]{stefanski1986optimally}
	Leonard~A Stefanski, Raymond~J Carroll, and David Ruppert.
	\newblock Optimally hounded score functions for generalized linear models with
	applications to logistic regression.
	\newblock \emph{Biometrika}, 73\penalty0 (2):\penalty0 413--424, 1986.
	
	\bibitem[Trapnell et~al.(2010)Trapnell, Williams, Pertea, Mortazavi, Kwan, van
	Baren, Salzberg, Wold, and Pachter]{trapnell2010transcript}
	Cole Trapnell, Brian~A Williams, Geo Pertea, Ali Mortazavi, Gordon Kwan,
	Marijke~J van Baren, Steven~L Salzberg, Barbara~J Wold, and Lior Pachter.
	\newblock Transcript assembly and quantification by rna-seq reveals unannotated
	transcripts and isoform switching during cell differentiation.
	\newblock \emph{Nature biotechnology}, 28\penalty0 (5):\penalty0 511--515,
	2010.
	
	\bibitem[Van De~Wiel et~al.(2012)Van De~Wiel, Leday, Pardo, Rue, Van Der~Vaart,
	and Van~Wieringen]{ShrinkBayes}
	Mark~A Van De~Wiel, Gwena{\"e}l~GR Leday, Luba Pardo, H{\aa}vard Rue, Aad~W Van
	Der~Vaart, and Wessel~N Van~Wieringen.
	\newblock Bayesian analysis of rna sequencing data by estimating multiple
	shrinkage priors.
	\newblock \emph{Biostatistics}, page kxs031, 2012.
	
	\bibitem[Wu et~al.(2011)Wu, Wang, and Zhang]{wu2011using}
	Zhengpeng Wu, Xi~Wang, and Xuegong Zhang.
	\newblock Using non-uniform read distribution models to improve isoform
	expression inference in rna-seq.
	\newblock \emph{Bioinformatics}, 27\penalty0 (4):\penalty0 502--508, 2011.
	
	\bibitem[Zhou et~al.(2014)Zhou, Lindsay, and Robinson]{edgeRrobust}
	Xiaobei Zhou, Helen Lindsay, and Mark~D Robinson.
	\newblock Robustly detecting differential expression in rna sequencing data
	using observation weights.
	\newblock \emph{Nucleic acids research}, page gku310, 2014.
	
	\bibitem[Zhu et~al.(1997)Zhu, Byrd, Lu, and Nocedal]{L_BFGS_B2}
	Ciyou Zhu, Richard~H Byrd, Peihuang Lu, and Jorge Nocedal.
	\newblock Algorithm 778: L-bfgs-b: Fortran subroutines for large-scale
	bound-constrained optimization.
	\newblock \emph{ACM Transactions on Mathematical Software (TOMS)}, 23\penalty0
	(4):\penalty0 550--560, 1997.
	
\end{thebibliography}

\end{document}